\documentclass[12pt]{article}
\usepackage{amsmath}
\usepackage{graphicx}
\usepackage{natbib}
\usepackage{url} 
\RequirePackage{amsthm,amsmath,amsfonts,amssymb}

\RequirePackage{graphicx}

\usepackage{algorithm,algpseudocode,latexsym,epsfig,amscd,multirow,paralist,dsfont}
\usepackage{latexsym,epsfig,graphicx,amsmath,amssymb,amscd,multirow,paralist,dsfont}

\usepackage[american]{babel}

\usepackage{epstopdf}

\usepackage{mathtools}
\usepackage{relsize}
\usepackage{graphicx}
\usepackage{caption}
\usepackage{subcaption}
\usepackage{hhline}
\usepackage[titletoc]{appendix}
\usepackage{array}
\newcolumntype{P}[1]{>{\centering\arraybackslash}p{#1}}
\usepackage[table]{xcolor}
\usepackage{siunitx,booktabs}
\usepackage{diagbox}

\newcommand{\thetavec}{{\boldsymbol{\theta}}}
\newcommand{\muvec}{{\boldsymbol{\mu}}}
\newcommand{\nuvec}{{\boldsymbol{\nu}}}

\newcommand{\bvec}{{\boldsymbol{b}}}
\newcommand{\zvec}{{\boldsymbol{z}}}

\newcommand{\zerovec}{{\boldsymbol{0}}}

\newcommand{\muhat}{\widehat{\mu}}
\newcommand{\muvechat}{\widehat{\muvec}}
\newcommand{\pihat}{\widehat{\pi}}
\newcommand{\tauhat}{\widehat{\tau}}
\newcommand{\mhat}{\widehat{m}}
\newcommand{\sigmahat}{\widehat{\sigma}}
\newcommand{\Sigmahat}{\widehat{\Sigma}}
\newcommand{\thetavechat}{\widehat{\thetavec}}

\newcommand{\diag}{\text{Diag}}
\newcommand{\trace}{\text{tr}}
\newcommand{\wh}{\widehat}

\newcommand{\nnoisy}{p_{\text{n}}}
\newcommand{\nsignal}{p_{\text{s}}}
\newcommand{\nbases}{h}
\newcommand{\npc}{q_{\text{c}}}
\newcommand{\lpen}{l_{\text{P}}}
\newcommand{\sign}{\textrm{sign}}
\newcommand{\signal}{\textrm{s}}
\newcommand{\noisy}{\textrm{n}}
\newcommand{\MAE}{\textrm{MAE}}
\newcommand{\BIC}{\textrm{BIC}}
\newcommand{\edf}{d_{\textrm{e}}}

\DeclareMathOperator*{\argmax}{argmax}

\begin{document}

	\def\spacingset#1{\renewcommand{\baselinestretch}%
		{#1}\small\normalsize} \spacingset{1}

	
	\title{Multivariate Functional Clustering with Variable Selection and Application to Sensor Data\\ from Engineering Systems}
	
	\author{
		Zhongnan Jin$^1$, Jie Min$^1$, Yili Hong$^1$, Pang Du$^1$, and Qingyu Yang$^2$ \textsf{}\\[1.5ex]
		{\small $^1$Department of Statistics, Virginia Tech, Blacksburg, VA}\\
		{\small $^2$Department of Industrial and Systems Engineering, Wayne State University, Detroit, MI}
	}
	
	\date{}
	
	\maketitle
	\bigskip
	\begin{abstract}
		Multi-sensor data that track system operating behaviors are widely available nowadays from various engineering systems. Measurements from each sensor over time form a curve and can be viewed as functional data. Clustering of these multivariate functional curves is important for studying the operating patterns of systems. One complication in such applications is the possible presence of sensors whose data do not contain relevant information. Hence it is desirable for the clustering method to equip with an automatic sensor selection procedure. Motivated by a real engineering application, we propose a functional data clustering method that simultaneously removes noninformative sensors and groups functional curves into clusters using informative sensors. Functional principal component analysis is used to transform multivariate functional data into a coefficient matrix for data reduction. We then model the transformed data by a Gaussian mixture distribution to perform model-based clustering with variable selection. Three types of penalties, the individual, variable, and group penalties, are considered to achieve automatic variable selection. Extensive simulations are conducted to assess the clustering and variable selection performance of the proposed methods. The application of the proposed methods to an engineering system with multiple sensors shows the promise of the methods and reveals interesting patterns in the sensor data. 
	\end{abstract}
	
	\noindent%
	\vfill
	
	\newpage
	\spacingset{1.3} 
	\section{Introduction}
	\label{sec:introduction}

	With the development of sensor and communication technologies, multi-sensor systems are now commonly deployed in many engineering systems to track both the system operating conditions and the system environment in a dynamic way. For example, sensors in aerospace play a vital role in navigation, detection, monitoring, and control of various air vehicles. Different types of aerospace sensors are located in the flight systems to measure the system states and vehicular environment (e.g., \citealp{nebylov2012}). Another example of multivariate sensor data is from automobiles, where multiple automotive sensors actively gather system and environment information, such as position, pressure, torque, exhaust temperature, and angular rate, from the powertrain, chassis, and body (e.g., \citealp{fleming2001}). In recent years, autonomous vehicles are targeting to operate without human involvement. To enable self driving, a variety of sensors, such as radar, lidar, sonar, GPS, and odometry, are deployed in autonomous vehicles to perceive their surroundings (e.g., \citealp{kocic2018}). In modern manufacturing systems, multiple sensors are installed to collect data (e.g., temperature, force, electrical signals, and vibration) from the manufacturing process, which can be utilized for different purposes such as the condition monitoring of machines, processes and tools, quality control, and resource management (e.g., \citealp{lee2013}).

Thus, sensor data are common in modern engineering systems and offer tremendous opportunities for statistical research to develop new methods for analyzing such data. From a business analytics and decision-making point of view, such new methods are much needed, for example, to answer questions such as, which sensor is important and what patterns there are in the sensor data. In the following, we describe one clustering application of sensor data to explore patterns in the data, which will be useful for exacting information from sensor data.

	Sensor data are generally collected at a high rate (e.g., per second). Thus, it is reasonable to treat the measurements from each sensor during a certain period as the observation of a functional variable. Then, a system with multiple sensors correspondingly produces multivariate functional data. During the system operation, a certain type of event can occur over time (e.g., the start/stop of the system, system faults, and breakdowns). It is insightful to investigate the sensor data in a window prior to the occurrence of the event. For each occurrence of the event, we can extract one multivariate functional observation from the sensor data stream. One practical interest is to cluster these multivariate functional observations to study the system operating patterns (i.e., different patterns in the occurrences of the event).  However, not all sensors contribute to the clustering. That is, some sensors in the system are irrelevant to the clustering. Thus, sensor selection is necessary.	
	
\begin{figure}
		\begin{center}
			\begin{tabular}{cc}
				\includegraphics[width=0.45\textwidth]{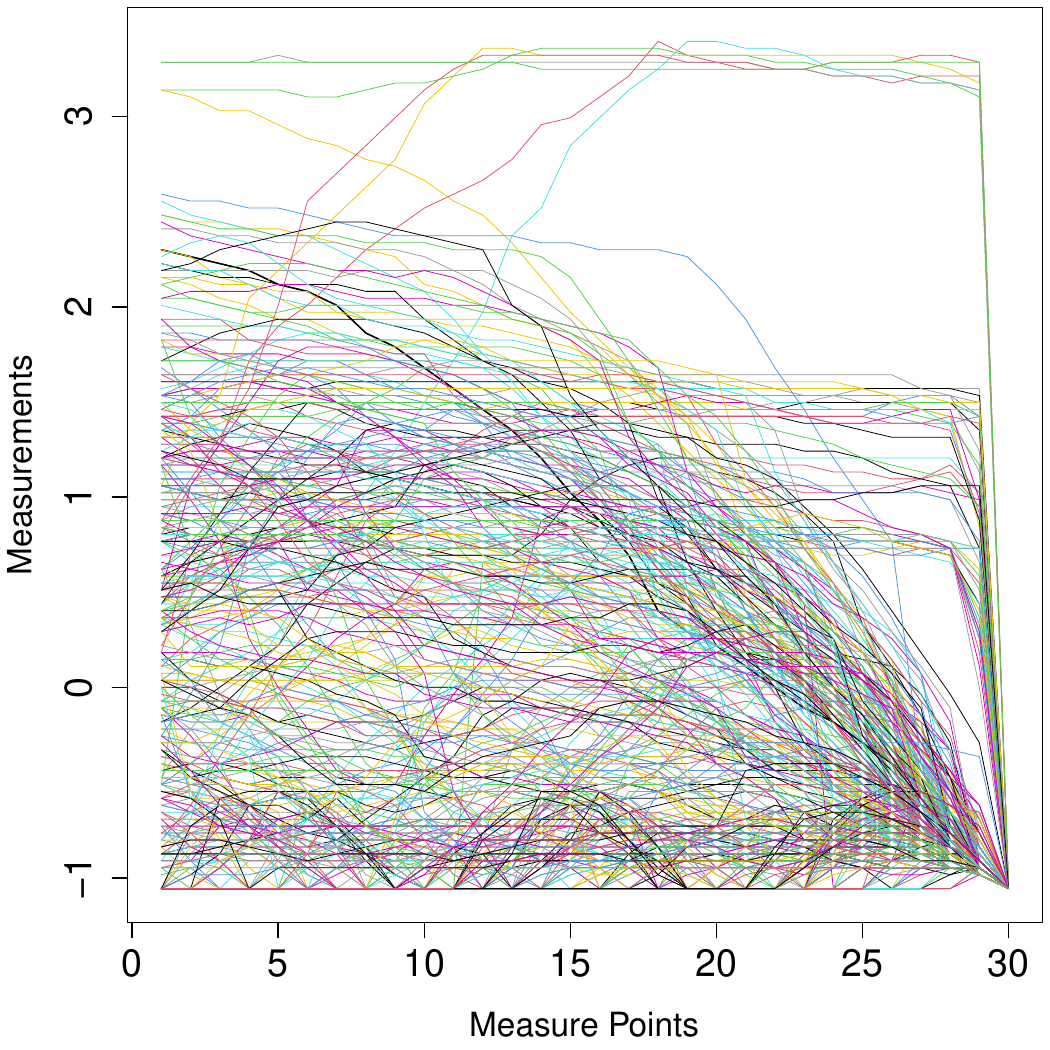}&
				\includegraphics[width=0.45\textwidth]{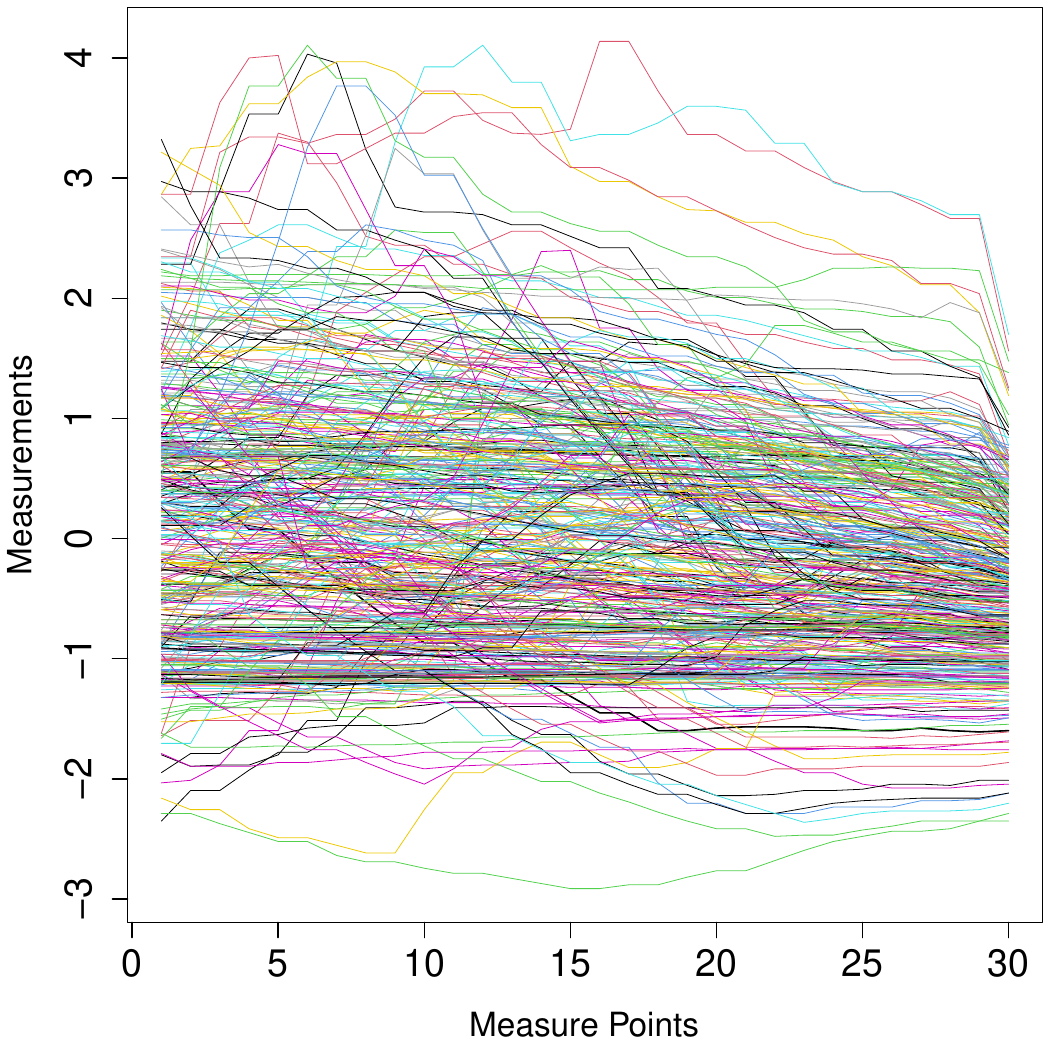}\\
				(a) Sensor 1 & (b) Sensor 2\\[2ex]
				\includegraphics[width=0.45\textwidth]{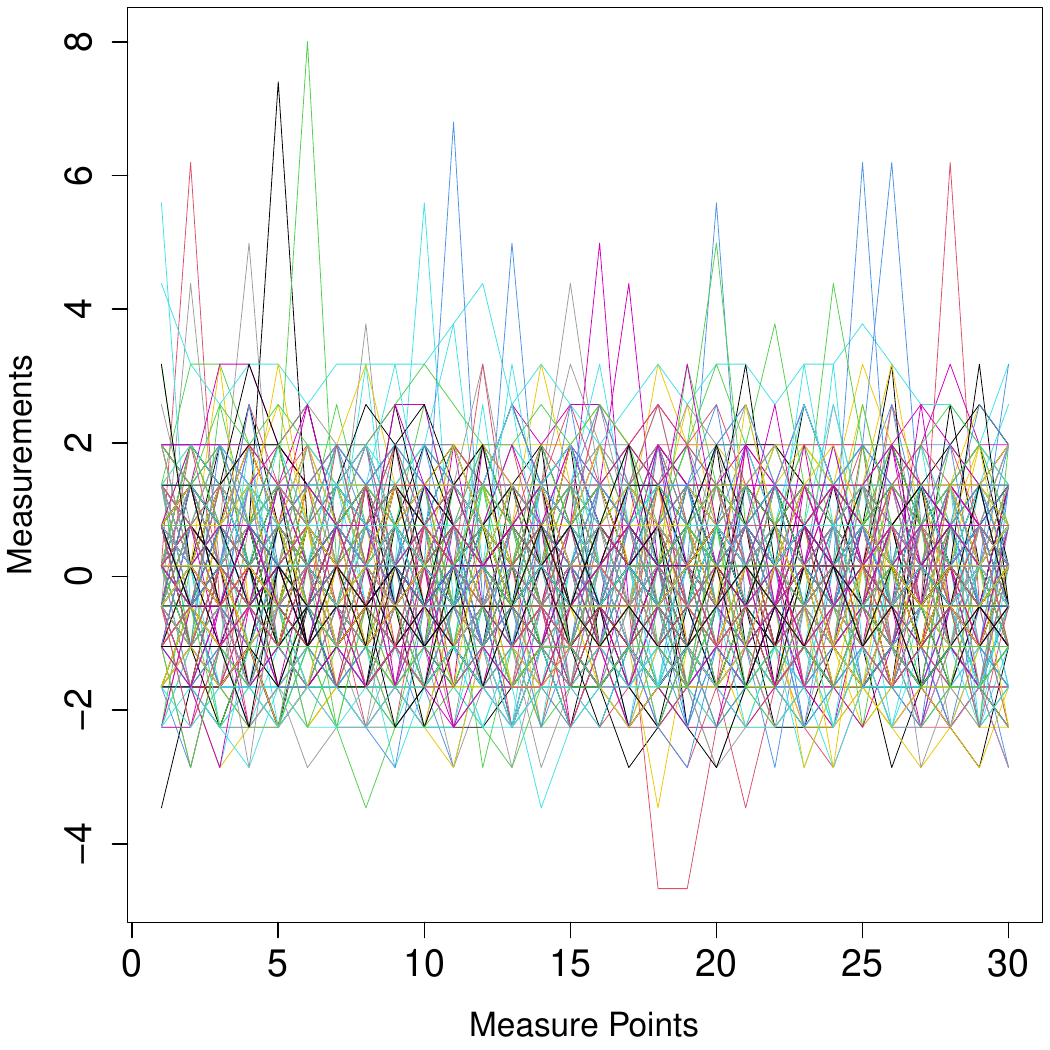} & \includegraphics[width=0.45\textwidth]{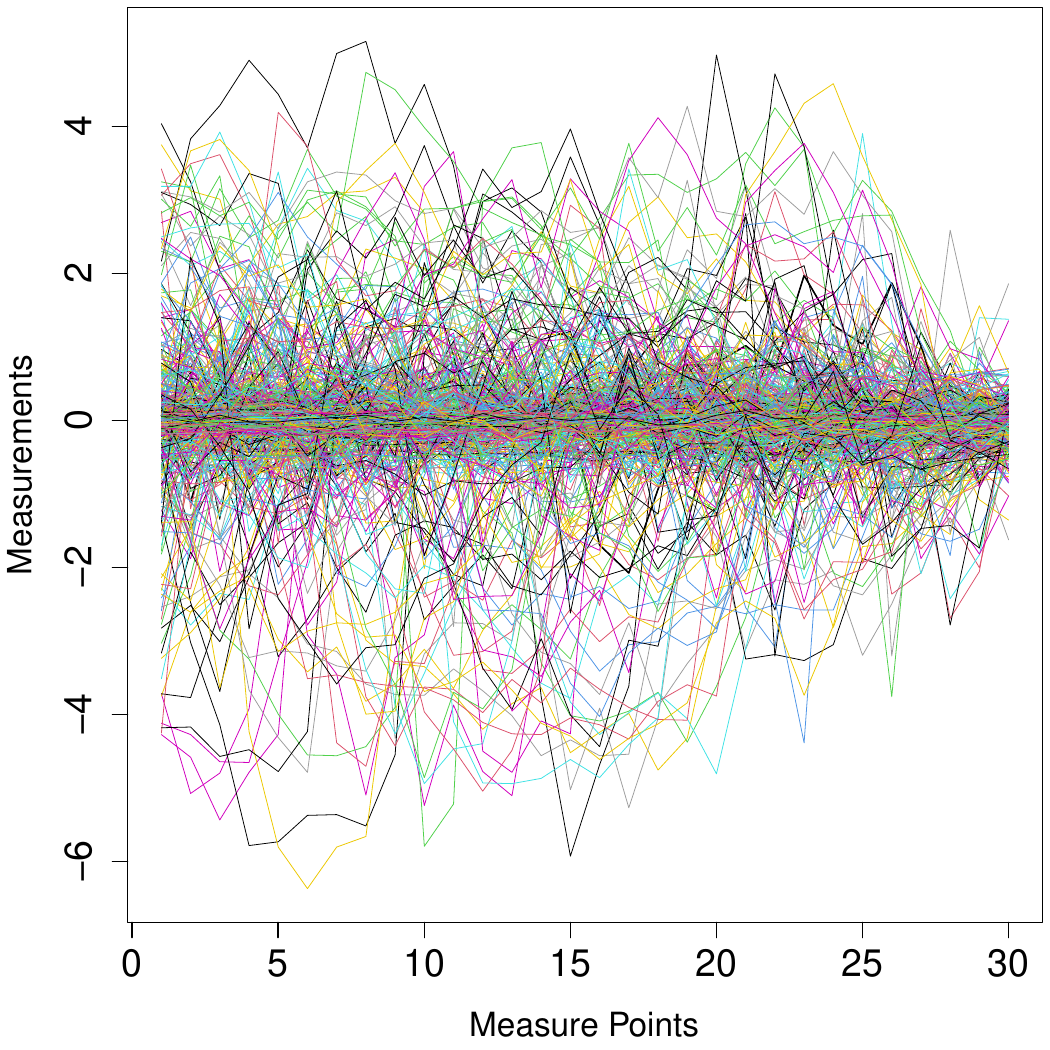}\\
				(c) Sensor 3 & (d) Sensor 4
			\end{tabular}
			\caption{Examples of engineering sensor data. Measurements from four sensors are shown. Each colored line shows one observed functional curve from one sensor.}\label{fig:engineerSensorExample}
		\end{center}
	\end{figure}

	This paper is motivated by a real engineering system application, which we call ``System~B.'' To protect proprietary sensitive information, we had to disguise the system and sensor names and use re-scaled data. In this system, 42 sensors are installed at various locations to monitor the operating status of the system. The sampling frequency of the sensors was 2 times per second at a regular grid. Figure~\ref{fig:engineerSensorExample} plots a subset of sensor data collected from System~B.  During the operation of the system, an event can occur due to various reasons, which is referred to as ``Event~S.'' To investigate the occurrence patterns of Event~S, engineers are particularly interested in 30 time points prior to the event occurrence (approximately 15 seconds). From the historical sensor data stream, there were 419 Event~S occurrences. For each occurrence, we extract one multivariate functional observation over the window that was 30 time points prior to the event occurrence. Figure~\ref{fig:engineerSensorExample} shows the 30-time point data collected before the 419 event occurrences from 4 of the 42 sensors. The goal was to find out the event occurrence patterns through the clustering of these 419 observations as well as selecting the sensors that were important contributors to these event occurrences. Therefore, we propose a clustering method for multivariate functional data with automatic variable selection.

	Regarding literature for functional data with clustering, \citet{abra03} applied the K-means algorithm to B-spline representations of functional data. \citet{james03} represented sparsely sampled functional data by spline basis functions and assumed Gaussian mixture distributions for the coefficient vectors with cluster-specific means. \citet{heard06} applied Bayesian hierarchical clustering to the basis expansion coefficients of curve data generated from a genetic study on malaria-infected mosquitoes. \citet{ray06} considered wavelet decomposition of functional data and then used a mixture of Dirichlet processes to specify prior beliefs about the regularity of the functions and the number of clusters. \citet{chiou07} and \citet{peng08} applied the K-means algorithm to functional data with distances respectively defined by the truncated Karhunen-Lo\`{e}ve expansion and the functional principal component scores. \citet{ma08} studied a mixed-effects functional ANOVA model where cluster probabilities for each curve can be estimated to provide a soft clustering of the curve. \citet{kayano10} expanded multivariate functional data by ortho-normalized Gaussian basis functions and then applied the self-organized map procedure to the resulting coefficients. \citet{serban12} considered clustering for multi-level functional data. They applied hard and soft clustering methods to scores derived from the multi-level functional principal component analysis (FPCA). \citet{chiou2012dynamical} proposed to use a multi-class logit model to estimate the posterior probability that an observed curve is in one cluster. \citet{gia13} used wavelet decomposition with thresholding to reduce the functional clustering problem to a linear mixed-effects model with unknown label variables and proposed an expectation-maximization (EM) algorithm to estimate the parameters. \citet{jacques2014model} proposed using multivariate functional principal component analysis (MFPCA) for dimension reduction, assumed a mixture normal distribution for principal component scores, and used EM algorithm in estimation. \citet{rodriguez2014functional} proposed to use splines and nested Dirichlet priors on spline coefficients in clustering Gaussian functional data where multiple curves were observed for each subject. \citet{jac14} provided a comprehensive survey on functional clustering methods. \citet{linton16} modeled each curve by the Nadaraya-Watson estimator and then applied an iterative thresholding procedure to order pairwise $L_2$ distances between curves to determine the number of clusters as well as cluster labels for individual curves. \citet{chamroukhi2019model} introduced a model-based classification model for functional data.
	
	More recently, \citet{jadhav2021pan} used B-splines to estimate the function curves. Through penalizing all pairwise spline coefficients, they combined estimation of coefficients and clustering together. \citet{zhong2021cluster} proposed to cluster non-Gaussian curves using a nonparametric transformation function with no prior information on the number of clusters and estimated the transformation function, number of clusters, and the mean and covariance functions of clusters simultaneously using penalized EM algorithm. \citet{guo2022functional} used multivariate state space model to represent the linear mixed effect model on functional data. Posterior probability is used to classify observations into different groups. Despite the rich literature on univariate functional clustering and multivariate functional clustering, none of the existing methods considered clustering and variable selection simultaneously.

	In this paper, we propose a penalized likelihood approach to perform simultaneous clustering and variable selection on multivariate functional data. FPCA is applied to the observations of each functional variable to represent them by linear combinations of leading functional principal components. The coefficients of these representations are then modeled through a Gaussian mixture distribution to achieve clustering. Group lasso type of penalties with adaptive weights are used for variable selection. Three penalties, namely, the individual, variable, and group penalties, are considered, each penalizing the mean components of the combined coefficients in a different way. Specifically, the individual penalty penalizes individual mean components so that each component is kept in or removed from the model by itself. The variable penalty penalizes the (absolute) maximum of all the mean components belonging to the same functional variable so that all these components are kept in or removed from the model simultaneously. The group penalty assumes a natural group of the functional variables and penalizes the mean components belonging to all the variables in the same group simultaneously so that all the mean components in this group are kept in or removed from the model together.

	To estimate the parameters, a complete log-likelihood is first formed with the inclusion of the latent cluster indicators and then one of the three penalties is added to form the penalized complete likelihood objective function. An EM algorithm is developed for the optimization, where closed-form expressions for parameter updates for each penalty are derived. The adjusted Bayesian information criterion (BIC) introduced by \citet{XiePanShen2008} is used for selecting tuning parameters such as the number of clusters, the penalty parameter, and the power of the adaptive weights in the penalty. Extensive simulations are conducted to assess the clustering and variable selection performance of the methods in various settings of sample sizes, signal-noise ratios, and signal strengths. The application of the proposed methods to the System~B data demonstrates that the proposed methods can reduce the number of sensors as well as reveal interesting patterns in the clustered event occurrences. Our method is the first functional clustering procedure that also incorporates variable selection.
	
	The rest of the paper is organized as follows.  Section~\ref{sec:proposed.method} describes the details of the developed methods. Performances of the proposed methods are evaluated by a simulation study in Section~\ref{sec:simulation.study}. The engineering system sensor data are analyzed in Section~\ref{sec:data.application}. Conclusion and remarks are included in Section~\ref{sec:conclusionRemark}.

	\section{Multivariate Functional Clustering with Automatic Variable Selection}\label{sec:proposed.method}
	
	We first introduce some notation. Let $p$ be the number of sensors (i.e., functional variables) and $n$ be the number of observations. The functional curve observed from the $s$th sensor of the $i$th observation is denoted by $x_{is}(t), i=1, \dots, n, s=1, \dots, p,  t\in [0, \tau]$ which is the observation period. In our engineering system study, $p=42$, $n=419$, and $\tau=30$. That is, we observe the values of  $x_{is}(t_j)$ for time points $0 < t_1 < \cdots < t_j < \cdots < t_{\tau}$. Here, $\tau$ is the number of time points.

	In the following, we introduce the proposed clustering method, which consists of two steps: (1) use FPCA to transform multivariate functional data to a coefficient matrix for data reduction; and (2) employ a model-based algorithm with penalty terms to perform clustering and automatic variable selection.

	\subsection{Functional Principal Component Transformation}\label{sec:MFPCA}
	In this section, we describe the details of the FPCA transformation. We focus on the transformation for observations on a single sensor. Let $\mu_{s}(t)$ be the mean function of $x_{is}(t)$. Let $\lambda_{s1}\ge\dots\ge \lambda_{sl}\ge \dots\ge 0$ and $\xi_{sl}(t)$ be the eigenvalues and eigenfunctions for $x_{is}(t)$, respectively. By the Karhunen-Lo\`{e}ve expansion, we have
	\begin{align} \label{sec:decomposition}
		x_{is}(t)=  \mu_{s}(t)+\sum _{l=1}^{\infty } c_{isl} \xi_{sl}(t) ,\quad t\in [0,\tau],
	\end{align}
	where $c_{isl}= \int_{0}^{\tau} [x_{is}(t) -  \mu_{s}(t)] \xi_{sl}(t) dt$ are random coefficients with mean 0 and variance $\lambda_{sl}$. A truncation of the expansion in \eqref{sec:decomposition} at a certain point would yield an approximate finite dimensional representation of the functional observation $x_{is}(t)$.

	In practice, the expansion in \eqref{sec:decomposition} can be estimated through the FPCA (e.g., \citealt{fda97}). After applying the FPCA to the observed functional curves from sensor $s$, we can obtain the functional principal component (FPC) functions $\widehat{\xi}_{sl}(t)$.  In practice, only the leading FPCs are kept, which can represent a relatively large portion of the variation in the original data. In this paper, we choose the first $\npc$ FPCs such that a relatively large portion of the variation for data from sensor $s$ is preserved. We will discuss the selection of the number of components $\npc$ in Section~\ref{sec: hyperparameter}.
	
	Let $\wh{c}_{isl}, l=1,\dots,\npc$ be the coefficients in the truncated expansion of $x_{is}(t)-\widehat{\mu}_s(t)$, where $\widehat{\mu}_s(t)$ is the empirical mean function for sensor $s$. We have,
	\begin{align}\label{equ:TruncatedKL}
		x_{is}(t)\approx  \widehat{\mu}_{s}(t)+\sum_{l=1}^{\npc} \wh{c}_{isl} \widehat{\xi}_{sl}(t),\quad t\in [0, \tau].
	\end{align}
	For notation simplicity, we let $b_{isl}=\wh{c}_{isl}$.  After the FPCA transformation is applied to the observations from all the sensors, we assemble all the coefficient vectors as follows,
	\begin{align}\label{eqn:y_isl}
		\bvec_i & =(b_{i11}, \dots, b_{i1\npc},\dots, b_{is1}, \dots, b_{is\npc}, \dots, b_{ip1}, \dots, b_{ip\npc})'\\\nonumber
		&=(\bvec_{i1}', \dots, \bvec_{is}', \dots, \bvec_{ip}')', \quad i=1,\dots, n,
	\end{align}
	where $\bvec_{is} =(b_{is1}, \dots, b_{is\npc})'.$ Note that the length of the coefficient vector $\bvec_i$ is $q = p\npc$.

	The coefficient vectors, together with the FPC functions, now represent a reduction from the original multivariate functional observations. For notation convenience, we also represent $\bvec_{i}$ in \eqref{eqn:y_isl} as,
	$$\bvec_i=(b_{i1}, \dots, b_{ij}, \dots, b_{iq} )', i=1,\dots, n.$$
	Note that here we use index $j$ to represent the elements in $\bvec_i$.

	In this paper, we apply separate FPCA for each sensor for data reduction, instead of applying MFPCA, because of the uniqueness of our problem. In literature, MFPCA is used for functional clustering when the number of functional variables is small (i.e., $p$ is around 2 to 3). For example, the number of functional variables of the applications considered in \cite{jacques2014model} is $p=2$. In our application, we have $p=42$, which is much larger than those in existing applications. Using MFPCA will lead to a challenging covariance estimation problem under large $p$. Instead, we set a common set of B-spline basis functions across all $p$ variables when we do separate FPCA. The basis functions $\xi_{sl}(t)$'s are linear combinations of these common B-spline basis functions, which can preserve the dependence among different functional variables to some extent. Similar ideas are also used in literature (e.g., \citealp{Kowaletal2017}).

	\subsection{Model Based Clustering with Variable Selection}\label{Intro_Var_Selection}
	We introduce a model-based algorithm with a penalty term to cluster the $q$-dimension coefficient vectors $\bvec_i, i=1, \dots, n$ obtained from Section~\ref{sec:MFPCA}. The coefficient matrix is $(\bvec_1, \dots, \bvec_n)'$, which is an $n\times q$ matrix. For convenience, we shall also call each column of the coefficient matrix as a variable.

	To conduct model-based clustering, the distribution of the random vector $\bvec_i$ is modeled by Gaussian mixture distributions, which are widely used in literature. The probability density function (pdf) of $\bvec_i$ is,
	$$g(\bvec_{i})=\sum_{k=1}^{m} \pi_{k} f_{k}(\bvec_{i};\muvec_{k},\Sigma).$$
	Here, $m$ is the number of clusters, $\pi_{k}$ are proportions satisfying $\sum_{k=1}^{m} \pi_{k} =  1$, and $f_{k}(\bvec_i;\muvec_{k},\Sigma)$ is the normal pdf for the $k$th cluster with mean $\muvec_{k}$ and covariance matrix $\Sigma$. That is,
	$$
	f_{k}(\bvec_{i};\muvec_{k},\Sigma) =  \frac{1}{(2\pi)^{\frac{q}{2}} |\Sigma|^{\frac{1}{2}}} \exp \left[  - \frac{1}{2} (\bvec_{i} - \muvec_{k})' \Sigma^{-1} (\bvec_{i} - \muvec_{k})   \right],
	$$
	where
	$$\muvec_{k} = (\mu _{k1}, \dots,\mu_{kj}, \dots, \mu _{kq})',$$
	and $\Sigma$ is the $q\times q$ covariance matrix. In the subsequent development, we also use the index structure as in \eqref{eqn:y_isl} to represent the elements in $\muvec_k$. That is,
	\begin{align}\label{eqn:mu_ksl}
		\muvec_k & =(\mu_{k11}, \dots, \mu_{k1\npc},\dots, \mu_{ks1}, \dots, \mu_{ks\npc}, \dots, \mu_{kp1}, \dots, \mu_{kp\npc})'\\\nonumber
		& =(\muvec_{k1}', \dots, \muvec_{ks}', \dots, \muvec_{kp}')', \quad k=1,\dots, m,
	\end{align}
	where $\muvec_{ks}=(\mu_{ks1}, \dots, \mu_{ks\npc})'$. For a parsimonious model and robustness in estimation, we use a diagonal structure for $\Sigma$ and set it common for all the clusters. That is, $$\Sigma=\diag(\sigma_1^2, \dots, \sigma_j^2, \dots, \sigma_q^2).$$
Note that the $\Sigma$ does not represent the correlations from the functional data, but it is related to the correlation of the coefficients in the KL expansion. A diagonal structure of $\Sigma$ is commonly used in literature. In summary, the vector for unknown parameters is $\thetavec=(\muvec_1', \dots, \muvec_m', \pi_1,\dots,  \pi_{m-1}, \sigma_1^2, \dots, \sigma_q^2)'.$
In some non-variable-selection context, separate covariance matrices are used for each cluster (e.g., \citealp{Schmutzetal2020}). In the variable selection setting, the number of variables can be large. We found separate covariance matrices can lead to robustness issues in covariance estimation when the numbers of observations assigned to some clusters are small.

	Let $\delta_{ik} $ be the indicator for observation $i$ in cluster $k$  such that
	$\delta_{ik} =1$ if   $\bvec_{i}$ is from cluster $ k$ and 0 otherwise.
	Then the negative log-likelihood with the latent indicators $\delta_{ik}$ is
	\begin{align} \label{equ: likelihood_fun}
		l(\thetavec)=-\sum_{i=1}^{n}\sum_{k=1}^{m} \delta_{ik} \left\{ \log(\pi_{k}) +\log \left[f_{k}(\bvec_{i}; \muvec_{k}, \Sigma)\right] \right\}.
	\end{align}
	To perform variable selection simultaneously with clustering, we consider the penalized negative log-likelihood
	\begin{align} \label{equ: likelihood_fun_penalty}
		\lpen(\thetavec)=-\sum_{i=1}^{n}\sum_{k=1}^{m} \delta_{ik} \left\{ \log(\pi_{k}) +\log \left[f_{k}(\bvec_{i}; \muvec_{k}, \Sigma)\right] \right\}+p_{\lambda}(\thetavec).
	\end{align}

	The penalty term $p_{\lambda}(\thetavec)$ in \eqref{equ: likelihood_fun_penalty} can have different forms according to the need. In this paper, we consider three types of penalties: the individual, variable, and group penalties. The individual penalty penalizes each mean component $\mu_{kj}$ separately, enforcing sparsity on individual means. The variable penalty penalizes the largest mean component corresponding to each variable in all clusters (i.e., $\max_{k} (|\mu_{kj} |)$), enforcing variable-wise mean sparsity. Note that the transformed data $\bvec_i$ has $q$ variables. The group penalty penalizes mean components from the same group simultaneously. Here we treat variables coming from the same sensor as one group.
	
	Specifically, the individual penalty is defined as
	$$p_{\lambda}(\thetavec) = \lambda \sum_{j=1}^{q} \sum_{k=1}^{m} w_{kj} |\mu_{kj}|,$$
	with weights $w_{kj}$. The variable penalty is defined as
	$$p_{\lambda}(\thetavec) = \lambda \sum_{j=1}^{q} w_{k^{\ast}j} \max_{k} (|\mu_{kj} |)$$
	with weights $w_{k^{\ast}j}$, where $k^{\ast} = \argmax_{k} (|\mu_{kj} |)$. The group penalty is defined as
	$$p_{\lambda}(\thetavec) = \lambda \sum_{k=1}^{m} \sum_{s=1}^{p} w_{k}\sqrt{\npc}||\muvec_{ks}||,$$
	with weights $w_k$, and $\muvec_{ks}$ is defined in \eqref{eqn:mu_ksl}. Here $||\cdot||$ is the $L_2$ norm of a vector.
	
	In each penalty, the weights serve a role that is similar to those in the adaptive Lasso \citep{Zou2006}. Specifically, $\lambda$ controls at the universal level since it remains the same for all variables and clusters and the weights adjust the penalization adaptively. In the individual penalty, one can define the weights according to mean components for each variable and cluster as in \citet{Zou2006}. That is,
	\begin{align}\label{eqn:UncontrainedMu}
		w_{kj} = \frac{1}{|\widetilde{\mu}_{kj}|^{\gamma}},
	\end{align}
	and $\widetilde{\mu}_{kj}$ is estimated when $\gamma=0$, which means no adaptive weight. In the variable penalty, we use the weights $w_{kj}$ as in the individual penalty but take the $k$ to be $k^{\ast}=\argmax_{k} (|\mu_{kj} |)$ for the same $j$. In the group penalty, we define the weights as
	\begin{align}\label{eqn:UnconstrainedMuGroup}
		w_{k} = \frac{1}{||\widetilde{\muvec}_{k}||^{\gamma}}.
	\end{align}
	Again, the element of $\widetilde{\muvec}_{k}$, which is $\widetilde{\mu}_{kj}$, is estimated when $\gamma=0$. In both \eqref{eqn:UncontrainedMu} and \eqref{eqn:UnconstrainedMuGroup}, $\gamma$ is a hyper-parameter for the weights. The calculations of $\widetilde{\mu}_{kj}$ in \eqref{eqn:UncontrainedMu} and \eqref{eqn:UnconstrainedMuGroup} are shown in Appendix \ref{Appendix:UnconstrianedMu}. With $\lambda$ and $\gamma$, the penalties put more weights on variables with mean components close to zero. On the other hand, variables with large mean components are not easily removed because the weights are small as well as the penalty terms. The hyper-parameter $\gamma$ adjusts the weights for mean components from each cluster. We need to choose hyper-parameters $\lambda$ and $\gamma$ simultaneously, which will be discussed in Section \ref{sec: hyperparameter}.
	
	Individual, variable, and group penalties all remove variables not contributing to the clustering procedure. However, according to the way each penalty is built, they act differently when removing variables. We now introduce more details on the principles behind all penalties.
	
	The individual penalty term penalizes individual mean component $\mu_{kj}$ for each cluster and each variable. For example, $\mu_{kj}$ and $\mu_{k'j}$ for $ k \neq k'$ are two separate parameters in the estimation even though $\mu_{kj}$ and $\mu_{k'j}$ both correspond to the $j$th variable. In the individual penalty, we remove the $j$th variable from the clustering procedure when all the corresponding mean components for all $m$ clusters are shrunk to zero, that is,
	$\mu_{1j} = \mu_{2j} = \cdots= \mu_{mj} = 0.$ For variables with little contribution to clustering, their mean components will be small across all the $m$ clusters and will be removed automatically by the individual penalty.
	
	The variable penalty sets the mean components $\mu_{kj}$ of the $j$th variable to zero for all $k= 1,\dots,m$, if its maximum mean component among $m$ clusters is set to zero. As a result, the variable removal criterion is $ \max_{k} (|\mu_{kj} |) = 0.$
	Compared to the individual penalty, the variable penalty considers the mean components from the same variable simultaneously since they share similar variable information. Note that $ \max_{k} (|\mu_{kj} |) = 0$ is equivalent to $\mu_{1j} = \mu_{2j}=  \cdots = \mu_{mj}  = 0$. However, how the individual penalty and the variable penalty set $\mu_{kj}$'s to zeroes are different. Individual penalty employs penalization on mean component for each variable, while variable penalty only applies penalization on the maximum mean component among different variables. For the $j$th variable, it is likely that some of the $\mu_{kj}$'s are set to zero, but not all. In this situation, using the individual penalty, the variable will not be removed. However, using the variable penalty, the variable is removed if the $\max_{k} (|\mu_{kj} |)$ is set to be zero. More details are in Section~\ref{sec: Parameter Estimation}.
	
	For the group penalty, the removal criterion is when all the elements of $\muvec_{ks}$ are set to zero for $k = 1, \dots, m$. In such case, the $s$th sensor is removed. For the individual and variable penalties, the $s$th sensor is removed when all of its mean $\mu_{ksl}$ are set to zero under the indexing in \eqref{eqn:mu_ksl}.
	
	With properly selected $\lambda$ and $\gamma$, all three penalties can automatically set the mean components to zero for those variables without much contribution to the clustering. Closed-form parameter estimations exist for all penalty terms, as described in the next section.

	\subsection{Estimation Procedure}\label{sec: Parameter Estimation}
	In estimation, we maximizes the penalized negative log-likelihood. From a decision-making point of view, the utility function is $l_p(\thetavec)$ in \eqref{equ: likelihood_fun_penalty}.  Because \eqref{equ: likelihood_fun_penalty} contains latent cluster indicators $\delta_{ik}$, we use an expectation-maximization (EM) algorithm to estimate the parameters $\thetavec$. The parameter updating step varies for different penalties. In the following, we describe the EM algorithms for individual, variable, and group penalties. In estimation, we first use the EM algorithm described below with $\gamma=0$ to obtain an estimate of $\muvec_k$ as $\widetilde{\muvec}_k$, and then perform the EM again with non-zero $\gamma$ to obtain the final estimates.
	
	In general, the EM algorithm consists of an expectation step (E-step) and a maximization step (M-step). The E-step estimates the cluster indicators $\delta_{ik}$ and updates the proportions $\pi_k$ through conditional expectations. The M-step updates the distributional parameters $ \muvec_{k}$ and $\Sigma$ based on the estimated $\delta_{ik}$ and $\pi_k$ from the E-step. The EM algorithm used in our method is shown as follows.
	
	\paragraph*{(i) Parameter Initialization}
	Given the number of clusters $m$ and the coefficient vectors $\bvec_{i}$, we initialize parameters $\pihat^{(0)}_{k}$, $\muvechat^{(0)}_{k}$ and $\Sigmahat^{(0)}$ for $k = 1, \dots, m$. We use the estimates from the K-nearest-neighborhood method to initialize the EM algorithm.
	
	\paragraph{(ii) E-Step}  \label{para:Estep}
	For the $r$th EM iteration, the conditional expectations of the cluster indicators $\delta_{ik}$ are estimated by
	$$\tauhat^{(r+1)}_{ik} = \frac{\pihat^{(r)}_{k} f_{k}(\bvec_{i}; \muvechat^{(r)}_{k}, \Sigmahat^{(r)})}{\sum_{k=1}^{m}\pihat^{(r)}_{k} f_{k}(\bvec_{i}; \muvechat^{(r)}_{k}, \Sigmahat^{(r)}) }\,, $$
	where $\muvechat^{(r)}_{k}$ and $\Sigmahat^{(r)}$ are the estimates from the previous step. The proportions are updated as
	$\pihat^{(r+1)}_{k} = n^{-1}{\sum_{i = 1}^{n} \tauhat^{(r+1)}_{ik}}. $
	
	\paragraph{(iii) M-Step} \label{para:Mstep}
	With the updated $\tauhat^{(r+1)}_{ik}$ and $\pihat^{(r+1)}_{k}$, the distributional parameters can be updated. In particular, the elements of $\Sigma$ are updated as
	\begin{align}\label{eqn:EMupdateCov}
		\widehat{\sigma}^{2,(r+1)}_{j} = \sum_{k=1}^{m} \sum_{i = 1}^{n} \tauhat^{(r)}_{ik} (b_{ij} - \muhat^{(r)}_{kj})^{2}/n, \quad  j=1, \dots, q.
	\end{align}
	The derivation of \eqref{eqn:EMupdateCov} is in Appendix~\ref{Appendix:CovarianceMatrixCal}. Note that the estimation of $\Sigma$ is the same for all the three penalty terms because the penalty terms do not involve the covariance matrix. On the other hand, the estimation procedures for the mean components vary for different penalties.
	
	The M-steps for mean components under different penalties are as follows. We follow the framework in \cite{XiePanShen2008}, but tailored the formulas for our specific model formulation with adaptive weights.
	
	\subparagraph*{(1) Individual Penalty} With adaptive weights, the sufficient and necessary condition for $\muhat_{kj}^{(r+1)}$ to globally minimize function $\lpen(\thetavec)$ is
	\[
	\left\{
	\begin{array}{ll}
		\dfrac{\sum_{i = 1}^{n} \tauhat_{ik}^{(r)} b_{ij}}{\sum_{i = 1}^{n} \tauhat_{ik}^{(r)}} = \left(  \dfrac{\lambda w_{kj}}{\sum_{i = 1}^{n} \tauhat_{ik}} + |\muhat_{kj}^{(r+1)}|  \right) \sign\left(\muhat_{kj}^{(r+1)}\right), \textrm{iff } \muhat_{kj}^{(r+1)} \neq 0\\[3ex]
		\dfrac{| \sum_{i = 1}^{n} \tauhat_{ik}^{(r)} b_{ij}|}{\sigmahat_{j}^{2, (r)}w_{kj}} \leq \lambda,  \quad \textrm{if } \muhat_{kj}^{(r+1)} = 0\\
	\end{array}
	\right. .
	\]
	Here, ``iff'' means ``if and only if.''  With the weight term $w_{kj}$ as specified in \eqref{eqn:UncontrainedMu}, the mean component update is different for each variable and cluster. In the $r$th step, we update $\mu_{kj}$ by
	\begin{align}\label{meanupdateIndividual}
		\muhat^{(r+1)}_{kj} = \frac{\sum_{i = 1}^{n} \tauhat^{(r)}_{ik}b_{ij}}{\sum_{i = 1}^{n} \tauhat^{(r)}_{ik}} \left( 1 - \frac{\lambda w_{kj}\sigmahat^{2,(r)}_{j}}{ |\sum_{i = 1}^{n}\tauhat^{(r)}_{ik}b_{ij} |} \right)_{+},
	\end{align}
	where $(x)_{+}=x$ if $x>0$ and 0 otherwise.
	In \eqref{meanupdateIndividual}, both $\sigmahat^{2,(r)}_{j}$ and $\tauhat_{ik}^{(r)}$ are calculated from the previous step.

	\subparagraph*{(2) Variable Penalty} Let $k^{\ast (r)}=\argmax_{k} (|\muhat_{kj}^{(r)}|)$, and
	\begin{align}\label{eqn:mu.hat.tilde}
		\widetilde{\widehat \mu}_{kj}^{(r)} = \frac{\sum_{i=1}^n \tauhat_{ik}^{(r)} b_{ij}}{\sum_{i = 1}^n \tauhat_{ik}^{(r)}}.
	\end{align}
	We update the parameter as follows,
	\[
	\muhat_{kj}^{(r+1)}=\left\{
	\begin{array}{ll}
		\widetilde {\widehat \mu}_{kj}^{(r)}, \quad \textrm{when} \quad k \neq k^{\ast (r)}\\
		\sign(\widetilde {\widehat \mu}_{kj}^{(r)}) \left(|\widetilde {\widehat \mu}_{kj}^{(r)}| - \dfrac{\lambda w_{kj} \sigmahat^{2, (r)}_{j}}{\sum_{i=1}^{n}\tauhat_{ik}}  \right)_{+},  \quad \textrm{when} \quad k =k^{\ast (r)}\\
	\end{array}
	\right. .
	\]

	\subparagraph*{(3) Group Penalty} In the group penalty, variables are grouped by sensors. The covariance matrix can be written as $\Sigma=\diag(\Sigma_{1},\dots, \Sigma_{s},\dots,  \Sigma_{m})$, where $\Sigma_{s}$ is a $\npc \times \npc$ diagonal matrix. With adaptive weights, we have the sufficient and necessary condition for $\muvechat_{ks}^{(r+1)}$ being a unique minimizer of $\lpen(\thetavec)$ as
	\[
	\left\{
	\begin{array}{ll}
		(\Sigmahat_{s}^{(r)})^{-1} \left[ \sum_{i = 1}^{n} \tauhat_{ik}^{(r)} \bvec_{is} - (\sum_{i = 1}^{n} \tauhat_{ik}^{(r)})\muvechat_{ks}^{(r+1)} \right] = \lambda w_{k} \sqrt{\npc} \dfrac{\muvechat_{ks}^{(r+1)} }{||\muvechat_{ks}^{(r+1)} ||}, \textrm{iff } \muvechat_{ks}^{(r+1)} \neq \zerovec\\
		
		||\sum_{i=1}^{n} \tauhat_{ik}^{(r)} (\Sigmahat_{s}^{(r)})^{-1}\bvec_{is}|| \leq \lambda w_{k} \sqrt{\npc},  \quad \textrm{if }\muvechat_{ks}^{(r+1)}  = \zerovec
	\end{array}
	\right. ,
	\]
	where $\bvec_{is}$ is defined in \eqref{eqn:y_isl}. As a result, in the M-step for the group penalty, $\muvec_{ks}$ is updated as
	\begin{align}\label{meanupdateGroup}
		\widehat{\muvec}_{ks}^{(r+1)} = \left[ \sign \left( 1 - \frac{\lambda w_{k} \sqrt{\npc}}{|| \sum_{i = 1}^{n} \tauhat_{kj}  (\Sigmahat_{s}^{(r)})^{-1}\bvec_{is} ||}  \right) \right]_{+} B_{ks} \widetilde{\muvechat}_{ks}^{(r)},
	\end{align}
	where
	$$B_{ks} = \left( I + \frac{\lambda w_k\sqrt{\npc}}{\sum_{i = 1}^{n} \tauhat_{kj} ||\muvechat_{ks}^{(r)}||} \Sigmahat_{s}^{(r)} \right)^{-1}, $$
	and the element of $\widetilde{\muvechat}_{ks}^{(r)}$ is as in \eqref{eqn:mu.hat.tilde}. Here, $I$ is the identity matrix.

	\paragraph*{(iv) Convergence}
	We repeat the steps introduced in (ii) and (iii) until it converges. In this paper, we set the stopping criterion as when the parameter changes in two consecutive steps are negligible, that is,
	when $|| \widehat{\thetavec}^{(r+1)} - \widehat{\thetavec}^{(r)} || \leq 0.0001.$

	\subsection{Hyper-parameter Selection}\label{sec: hyperparameter}
     For each type of the penalties, it is important to select the value of hyper-parameters. There are three hyper-parameters $\lambda$, $\gamma$, and $m$. Recall that $\lambda$ and $\gamma$ are for the penalty terms, and $m$ is the number of clusters. Following \cite{XiePanShen2008}, we use the adjusted BIC to choose the hyper-parameters, which is  defined as,
	$
	\BIC = 2l(\thetavechat) + \log(n)\edf.
	$
	Here $\edf= m + p +mp - n_{0}-1 $ is the effective degrees of freedom and $n_{0}$ is the number of mean components that are forced to be zeros. Because we model all variables with a diagonal covariance structure, the information provided by data is not only related to the number of observations, but also related to the number of variables. Therefore, we slightly modify the calculation of adjusted BIC as,
	\begin{align}\label{eqn:AdjBIC}
		\BIC = 2l(\thetavechat) + \log(np)\edf.
	\end{align}
	
	In literature, choices of $\lambda$ and $m$ are well studied while the hyper-parameter $\gamma$ in the weights is often set to be 1. To have more flexible weights, we use a grid search to find the hyper-parameter combination with the minimum adjusted BIC. To avoid extreme behavior of weights, we fix the $\gamma$ grid as $(0.5,1,1.5,2)$.
	
	Because the best $\lambda$ value increases as the number of observations increases, we use different grids for different numbers of observations. In particular, the grid for $\lambda$ is $(0,0.5,1,2,3,5,7,10, 15,20) \times n^{1/3}$. To choose the best $\gamma$ and $\lambda$ combination, with each value of $\gamma$ in $\gamma$ grid, we first fix $\gamma$ at zero, use adjusted BIC to choose the $\lambda$, and use the chosen $\lambda$ to estimate $\widetilde{\muvec}_k$ which are used in the adaptive weights as in \eqref{eqn:UncontrainedMu} and \eqref{eqn:UnconstrainedMuGroup}. Then, we perform estimation again with this $\widetilde{\muvec}_k$ at the non-zero $\gamma$ we choose, and go through the $\lambda$ grid again using the adaptive Lasso. We use adjusted BIC to choose the best $\gamma$ and $\lambda$ combination.
	
	It is easy to see from \eqref{eqn:AdjBIC} that the adjusted BIC becomes smaller when more mean components are set to zeros. Since using more clusters and variables leads to a larger log-likelihood, the numbers of clusters and variables are controlled by the penalty term in \eqref{eqn:AdjBIC}. Therefore, the chosen hyper-parameters reach a balance between the goodness-of-fit represented by the log-likelihood and model size restricted by the penalty.

	Another specification that needs to be made is the number of FPCs, $\npc$, which is a non-trivial issue. One commonly used strategy is to select the number of FPCs so that a high percentage, say 90\%, of variation is explained by the FPCA, which works well when $p$ is small. When $p$ is large, requiring all functional variables achieve a high percentage of variation would require a large number of FPCs.  Another strategy is to choose different $\npc$ for each functional variable according to some model criterion (e.g., \citealp{zhong2021cluster}). Doing that will create unequal weights across different functional variables in the likelihood calculation (i.e., those functional variables with more FPCs will contribute more to the likelihood), which does not fit the clustering problem considered in this paper.
	In this paper, we use a simple but practical solution. We set the number of FPCs so that at least $(100\alpha)\%$ of the functional variables have more than $(100\beta)\%$ of their variations explained by the chosen number of FPCs. In the data analysis, we set proportions as $\alpha=0.8$ and $\beta=0.8$ to select the number of FPCs. That is, we select $\npc$ such that at least $80\%$ of the sensors have more than $80\%$ of their variations explained by their first $\npc$ FPCs.

To link to the decision-making framework, the above sections propose a functional data clustering method, provide criteria and suggestions in selecting the number of clusters, the penalty type, and the value of penalties. To obtain a satisfying clustering result, it is important to select the number of clusters, a proper type of penalty, and also the values of penalties.

	\section{Simulation Study}\label{sec:simulation.study}
	In this section, we conduct simulations on various scenarios to evaluate the performance of clustering and variable selections. We simulate data that mimic the real data from System~B with enough sampling points to represent the true functional curves of all the sensors. After generating data, we apply the FPCA, and conduct clustering and variable selection under the optimal hyper-parameters chosen by the adjusted BIC. Estimated results are compared with true values, and model performances with different penalties are discussed. To illustrate the advantage of variable selection in clustering, we also compare the results from clustering with variable selection with the results from clustering without penalty (i.e.,  no variable selection).

	\subsection{Setup}
	In the simulation, we investigate the effects of three factors on clustering and variable selection, which are the sample size, the signal-noise ratio, and the signal strength. The signal-noise ratio is reflected by the number of signal sensors and the number of noisy sensors. In particular, we consider both signal sensors (i.e., have effects on clustering) and noisy sensors (i.e., have no effects on clustering). Let $\nsignal$ be the number of signal sensors and $\nnoisy$ be the number of noisy sensors. The total number of sensors is $p=\nsignal+\nnoisy$. Signal strength refers to the variability of functional curves around its cluster center.
	
	To allow for investigating those factors in the simulation in a relatively simple manner, certain structures are used in the data simulations. For both signal sensors and noisy sensors, we use B-splines with $\nbases=12$ bases of order 3. The domains of all simulated sensor data are from 0 to 30, where there are 31 measurements on each unit. With the same basis functions,  $\nbases$ coefficients are used for each sensor.  The spline coefficients for an observation $i$ are $\zvec_i = [(\zvec_{i}^{\signal})', (\zvec_{i}^{\noisy})']'$ with elements,
	\begin{align*}
		\zvec_{i}^{\signal}&= \left(z_{i11}^{\signal}, \dots, z_{i1\nbases}^{\signal},\dots,z_{ i\nsignal1}^{\signal}, \dots,z_{i\nsignal\nbases}^{\signal}\right)',\\
		\zvec_{i}^{\noisy}&= \left(z_{i11}^{\noisy}, \dots, z_{i1\nbases}^{\noisy},\dots,z_{ i\nnoisy1}^{\noisy}, \dots,z_{i\nnoisy\nbases}^{\noisy}\right)'.
	\end{align*}
	Here, $\zvec_{i}^{\signal}$ and $\zvec_{i}^{\noisy}$ are spline coefficients for signal sensors and noisy sensors, respectively. To simulate multiple functional clusters, we use a Gaussian mixture distribution to generate $\zvec_i$, whose pdf is,
	\begin{align}\label{eqn:sim.model}
		g(\zvec_i ) = \sum_{k=1}^{m} \pi_{k} f_{k}(\zvec_i ;\nuvec_{k},\Upsilon_{k}).
	\end{align}
	Note that the simulation model in \eqref{eqn:sim.model} is similar to, but different from, the fitted model in \eqref{equ:TruncatedKL}. In this way, data simulated from \eqref{eqn:sim.model} can mimic the main features in the real data but are not necessarily in favor of the fitted model, because the simulated data still need to go through the FPCA for data reduction. Also, before performing the analysis, we re-scale the data of each sensor to have mean $0$ and variance $1$.

\begin{figure}
		\begin{center}
			\begin{tabular}{ccc}
				\includegraphics[width=0.3\textwidth]{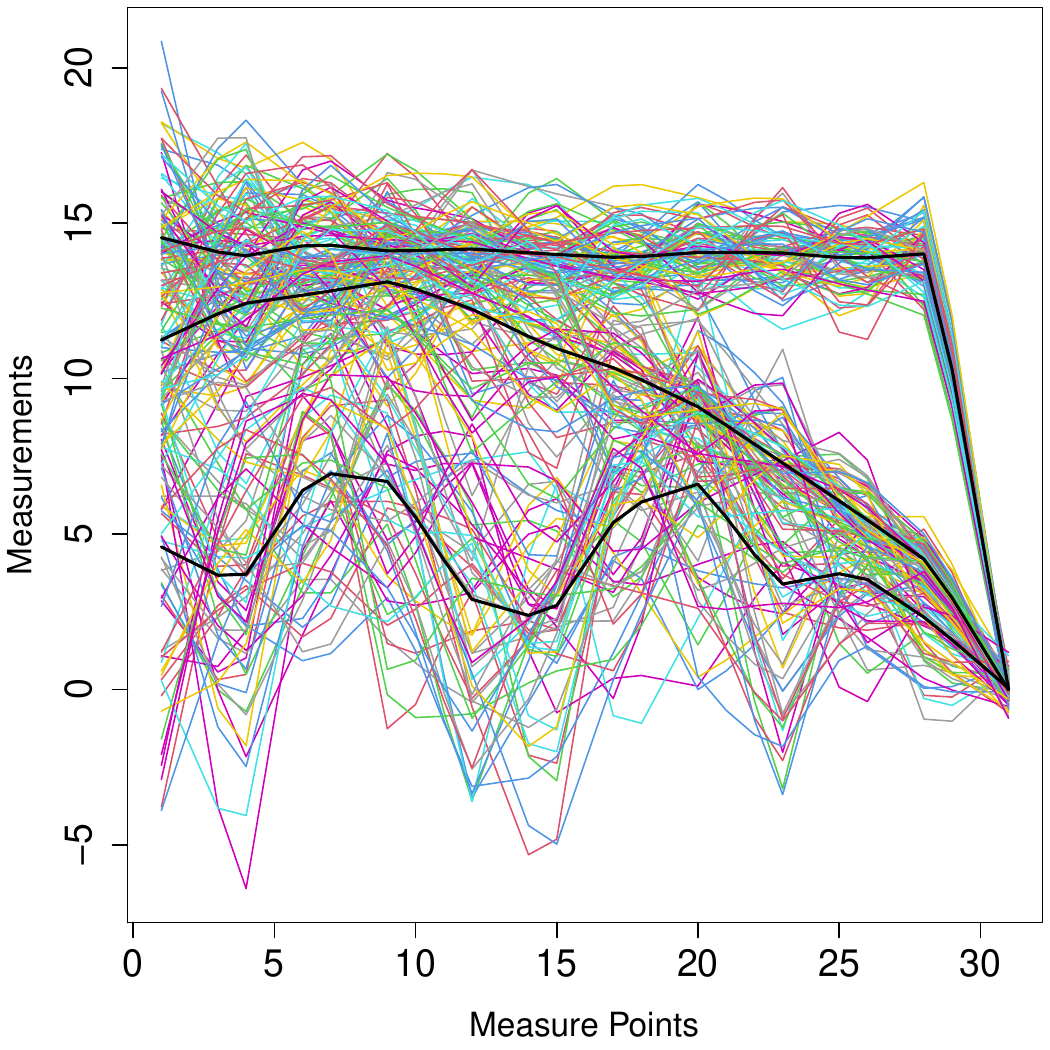}&
				\includegraphics[width=0.3\textwidth]{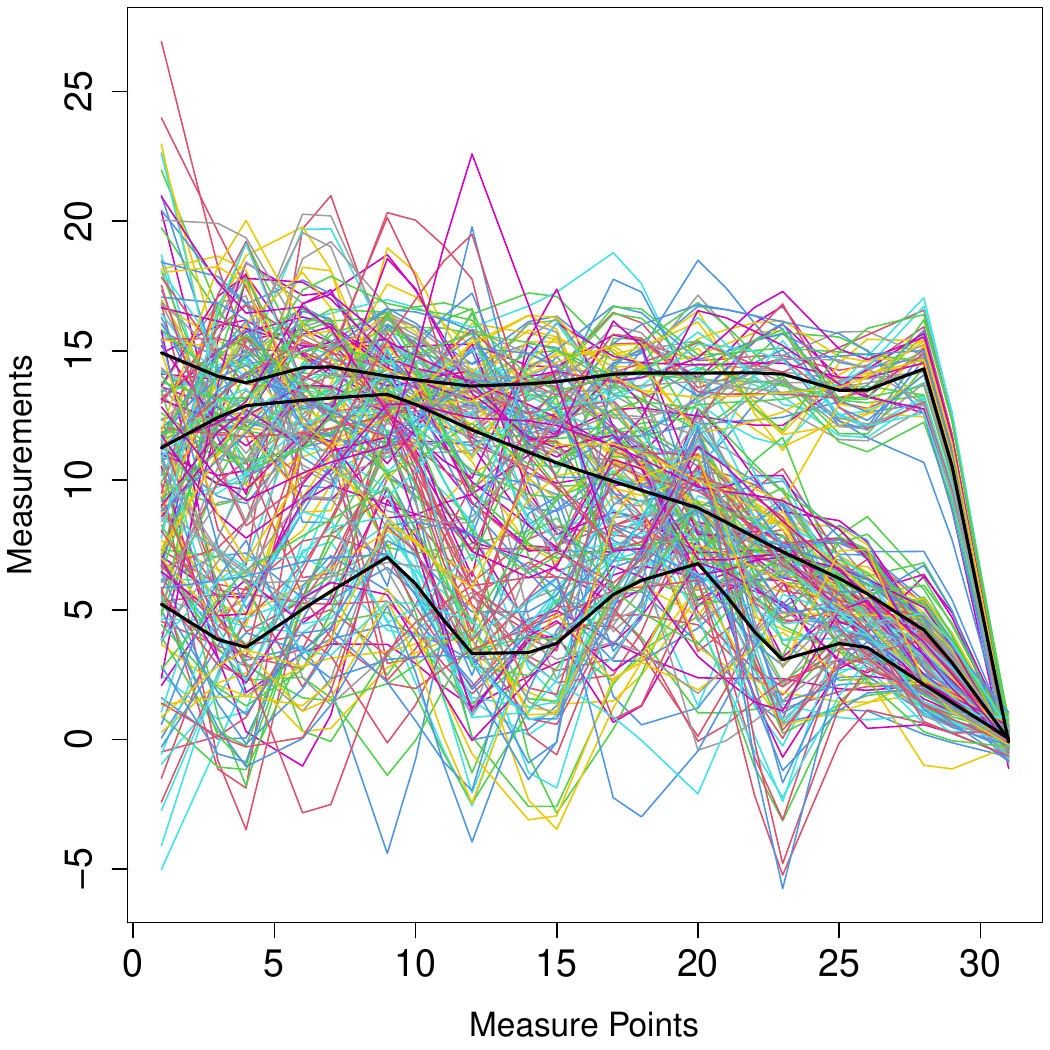}&
				\includegraphics[width=0.3\textwidth]{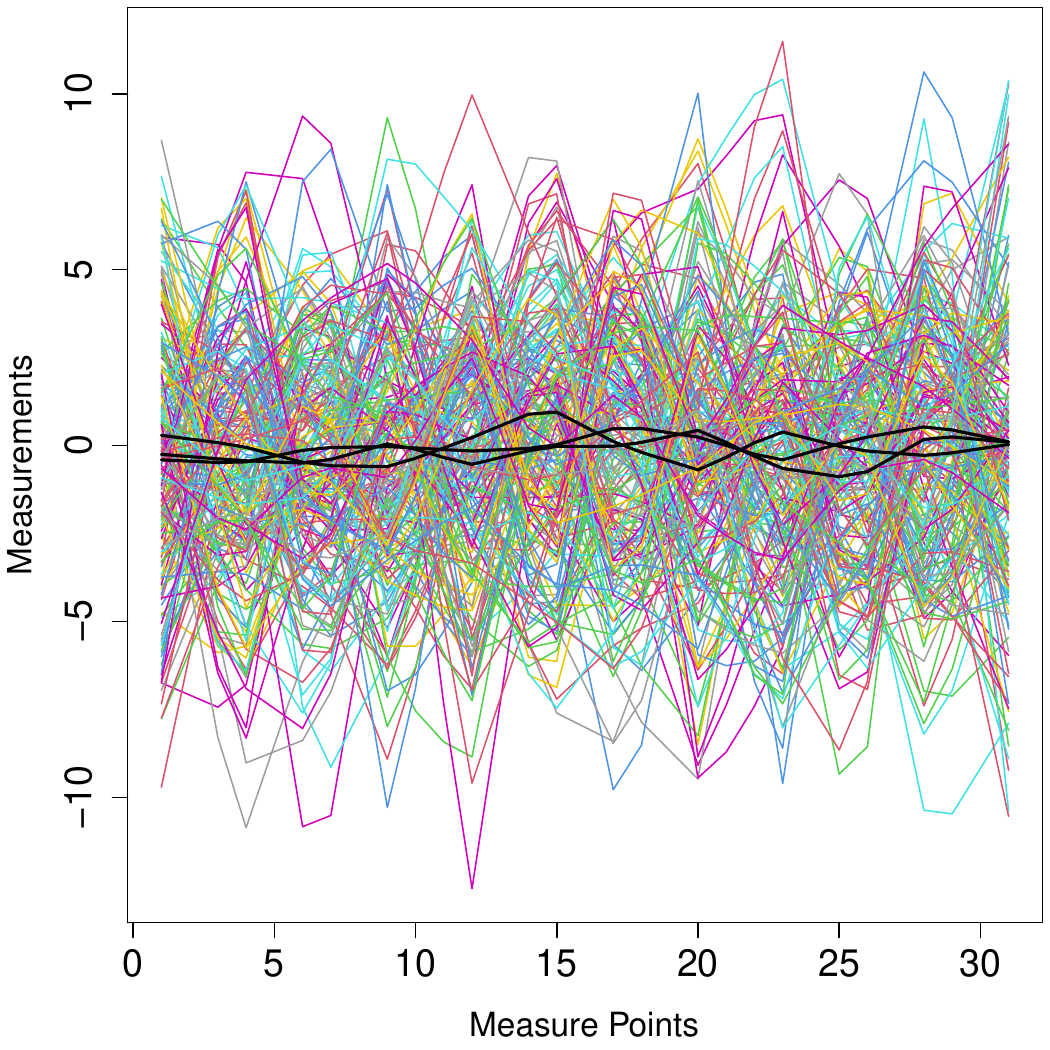}\\
				(a) Strong Signal Sensor & (b) Weak Signal Sensor & (c) Noisy Sensor
			\end{tabular}
			\caption{Examples of simulated functional curves from a strong signal sensor ($\delta=2.5$), a weak signal sensor ($\delta=1$), and a noisy sensor. Each colored line shows one observed functional curve from one sensor. The solid black lines represent cluster means.}\label{fig:simdat.simvar}
		\end{center}
		
	\end{figure}

	To simplify the setting, we set the number of clusters $m=3$ and use equal proportions with $\pi_{k} = 1/3$, $k = 1, 2, 3$. The mean component for cluster $k$ is $\nuvec_k = [(\nuvec_{k}^{\signal})', (\nuvec_{k}^{\noisy})']'$ with elements,
	\begin{align*}
		\nuvec_{k}^{\signal} &= (\nu_{k11}^{\signal}, \dots, \nu_{k1\nbases}^{\signal},\dots,\nu_{k\nsignal1}^{\signal}, \dots,\nu_{k\nsignal\nbases}^{\signal})',\\
		\nuvec_{k}^{\noisy} &= (\nu_{k11}^{\noisy}, \dots, \nu_{k1\nbases}^{\noisy},\dots,\nu_{k\nnoisy1}^{\noisy}, \dots,\nu_{k\nnoisy\nbases}^{\noisy})'.
	\end{align*}
	Here, $\nuvec_{i}^{\signal}$ and $\nuvec_{i}^{\noisy}$ are the mean components for signal sensors and noisy sensors, respectively. The covariance matrix for cluster $k$ is specified as $\Upsilon_k = \diag(\Upsilon_{k}^{\signal}, \Upsilon_{k}^{\noisy})/\delta$, and
	\begin{align*}
		\Upsilon_{k}^{\signal} &= \diag\left[(\upsilon_{k11}^{\signal})^2, \dots, (\upsilon_{k1\nbases}^{\signal})^2, \dots, (\upsilon_{k\nsignal1}^{\signal})^2, \dots, (\upsilon_{k\nsignal\nbases}^{\signal})^2\right],\\
		\Upsilon_{k}^{\noisy} &= \diag\left[(\upsilon_{k11}^{\noisy})^2, \dots, (\upsilon_{k1\nbases}^{\noisy})^2, \dots, (\upsilon_{k\nsignal1}^{\noisy})^2, \dots, (\upsilon_{k\nsignal\nbases}^{\noisy})^2\right].
	\end{align*}
	Here, $\delta$ is a value used to control the magnitude of variance (i.e., control the signal strength), and $\Upsilon_{k}^{\signal}$ and $\Upsilon_{k}^{\noisy}$ are the variance terms for signal sensors and noisy sensors, respectively. For signal sensors, we choose the value of $\nuvec_{k}^{\signal}$ and $\Upsilon_{k}^{\signal}$ to mimic the real data. For all $m$ clusters of noisy sensor data, we set elements in $\nuvec_{k}^{\noisy}$ to be all $0$, and assign the same value to all the diagonal elements in $\Upsilon_{k}^{\noisy}$. The value is chosen so that variances of spline coefficients for noisy sensors have a similar magnitude to the variances of spline coefficients for signal sensors. As an illustration, Figures~\ref{fig:simdat.simvar}(a), \ref{fig:simdat.simvar}(b), and \ref{fig:simdat.simvar}(c) show examples of simulated functional data from a strong signal sensor, a weak signal sensor, and a noisy sensor, respectively.

	In our simulations, we consider three scenarios, which investigate the effects of sample size, the signal-noise ratio, and the signal strength on clustering and variable selection. To see the effect of sample size, we vary $n = $50, 200, 350 and 500 to generate data, and fix $\nsignal = 2$, $ \nnoisy = 16 $ and $\delta = 1.5$.
	
	We study the effect of the signal-noise ratio by varying the number of noisy sensors while the number of signal sensors is fixed. We set $\nsignal = 2$ while using $\nnoisy = 8, 16, 32$ and $64$. We fix $n = 200$ and $\delta=1.5$ for all cases in this scenario.
	
	We investigate the effect of signal strength by varying $\delta=1, 1.5, 2$, and $2.5$. A larger $\delta$ leads to a smaller coefficient variance, which generates functional curves that are closer to each other. On the other hand, a smaller $\delta$ generates functional curves that are more spread out. We fix $\nsignal = 2, \nnoisy = 16$ and $n = 200$ for all the cases in this scenario.

	\subsection{Results}
	In this section, we present simulation results with different sample sizes, signal-noise ratios, and signal strengths. Under each setting, we repeat 200 times and obtain the results. For the 200 simulated samples under each setting, we calculate the mean absolute error of the estimated number of clusters as
	$\MAE(\mhat)=\sum_{i = 1}^{200}|\widehat{m}_i - m_{\textrm{True}}|/200,$
	where $\widehat{m}_i$ is the estimated number of clusters for the $i$th sample and $m_{\textrm{True}}$ is the true number of clusters. In this study, $m_{\textrm{True}}$ is 3 across different setups.
	For clustering accuracy, we also compute the adjusted Rand index (ARI), which is a nonparametric correlation estimation \citep{Hubert1985}, where a higher similarity between two variables returns an ARI closer to 1.
	
	To evaluate the performance of variable selection, we calculate the mean number of variables removed, sensors removed correctly, and sensors removed falsely. Signal sensors are considered as removed falsely if they are identified as noises. In the simulation, the number of FPCs is chosen as three (i.e., three variables for one sensor).
	
\begin{figure}
		\begin{center}
			\includegraphics[width=0.85\textwidth]{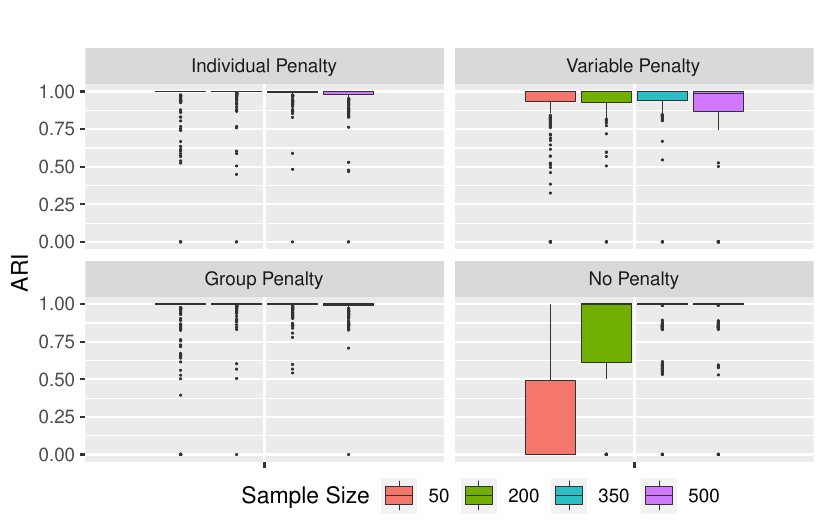}
			\caption{Boxplots of ARI for 200 datasets under different sample sizes for various methods.} \label{fig:simresult}
		\end{center}	
	\end{figure}

	Table~\ref{Table:SampleSize} shows the summary of simulation results, and Figure~\ref{fig:simresult} shows the boxplots of ARI for 200 datasets under different sample sizes for various methods. The $\MAE(\mhat)$'s using different penalties are similar. From the ARI plot, we can see that the performance of the individual penalty and the group penalty are similar, but the ARI of the variable penalty is worse than the individual and group penalties. Regarding variable selection, the number of falsely removed sensors is generally small. The variable penalty tends to have more falsely removed sensors compared with the individual and group penalties. The number of correctly removed sensors is close to the true value (i.e., 16) for all the cases and all the three different penalties.

	\begin{table}
		\caption{Summary of simulation results under $n=50, 200, 350$ and 500, with $\delta=1.5$, $\nsignal = 2$, and $\nnoisy = 16$ for all settings. }
		\label{Table:SampleSize}
		\centering
		\begin{tabular}{c|c|c|c|c|c||c}\hline\hline
			\multirow{2}{*}{$n$}& \multirow{2}{*}{Penalty} & \multirow{2}{*}{$\MAE(\mhat)$ }& Variable & Sensor removed & Sensor removed & $\MAE(\mhat)$    \\
			&  & &removed &  correctly& falsely & (no penalty) \\
			\hline
			\multirow{3}{*}{50} &
			Individual &
			0.29 & 49.88 & 15.84 & 0.05 &  \\
			&
			Variable &
			0.20 & 49.37 & 15.32 & 0.19 & 1.70 \\
			&
			Group &
			0.36 & 47.55 & 15.74 & 0.11 &  \\
			\hline
			\multirow{3}{*}{200} &
			Individual &
			0.26 & 49.95 & 15.91 & 0.02 &  \\
			&
			Variable &
			0.28 & 50.15 & 15.76 & 0.08 & 0.46 \\
			&
			Group &
			0.20 & 47.56 & 15.84 & 0.01 &  \\
			\hline
			\multirow{3}{*}{350} &
			Individual &
			0.32 & 49.50 & 15.82 & 0.01 &  \\
			&
			Variable &
			0.26 & 49.99 & 15.78 & 0.06 & 0.25 \\
			&
			Group &
			0.28 & 47.42 & 15.80 & 0.00 & \\
			\hline
			\multirow{3}{*}{500} &
			Individual &
			0.43 & 48.96 & 15.59 & 0.03 & \\
			&
			Variable &
			0.42 & 49.48 & 15.51 & 0.17 & 0.22 \\
			&
			Group &
			0.28 & 46.81 & 15.60 & 0.01 &  \\
			\hline\hline
		\end{tabular}
		
	\end{table}

\begin{figure}
		\begin{center}
			\includegraphics[width=0.85\textwidth]{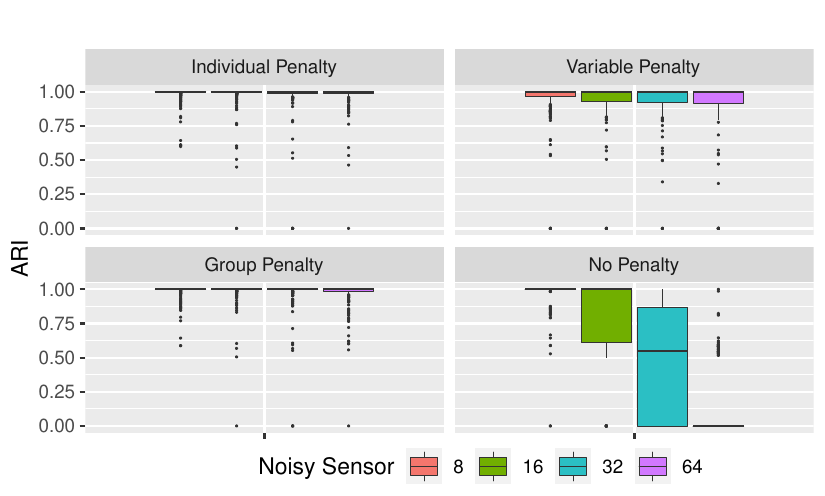}
			\caption{Boxplots of ARI for 200 datasets under different numbers of noisy sensors for various methods.} \label{fig:simresult.noisySensor}
		\end{center}	
	\end{figure}

	Table~\ref{Table:Noisyevaluation} shows the summary of simulation results, and Figure~\ref{fig:simresult.noisySensor} shows the boxplots of ARI for 200 datasets under different signal-noise ratios for various methods. The group penalty method performs better than the individual and variable methods in terms of $\MAE(\mhat)$ and the number of falsely removed sensors. Also, the number of correctly removed sensors is close to the true value for all the cases and all the three different penalty methods, and the variable penalty tends to have more falsely removed sensors. The ARI plots show that the individual and group penalties are better than the variable penalty.

	\begin{table}
		\caption{Summary of simulation results under $\nnoisy= 8, 16, 32,$ and $64$, with $n=200$, $\delta=1.5$, and $\nsignal = 2$ for all settings.}
		\label{Table:Noisyevaluation}
		\centering
		
		\begin{tabular}{c|c|c|c|c|c||c}\hline\hline
			\multirow{2}{*}{$\nnoisy$}& \multirow{2}{*}{Penalty} & \multirow{2}{*}{$\MAE(\mhat)$ }& Variable & Sensor removed & Sensor removed & $\MAE(\mhat)$    \\
			&  & &removed &  correctly& falsely & (no penalty) \\
			\hline
			\multirow{3}{*}{8} &
			Individual &
			0.28 & 25.96 & 7.93 & 0.01 &  \\
			&
			Variable &
			0.20 & 26.29 & 7.89 & 0.09 & 0.12 \\
			&
			Group &
			0.18 & 23.85 & 7.94 & 0.00 & \\
			\hline
			\multirow{3}{*}{16} &
			Individual &
			0.26 & 49.95 & 15.91 & 0.02 &  \\
			&
			Variable &
			0.28 & 50.15 & 15.76 & 0.08 & 0.46 \\
			&
			Group &
			0.20 & 47.56 & 15.84 & 0.01 & \\
			\hline
			\multirow{3}{*}{32} &
			Individual &
			0.34 & 98.12 & 31.98 & 0.06 & \\
			&
			Variable &
			0.27 & 97.94 & 31.67 & 0.10 & 1.17 \\
			&
			Group &
			0.26 & 95.64 & 31.84 & 0.04 &  \\
			\hline
			\multirow{3}{*}{64} &
			Individual &
			0.32 & 189.09 & 62.26 & 0.02 & \\
			&
			Variable &
			0.32 & 184.74 & 60.65 & 0.14 & 1.81 \\
			&
			Group &
			0.29 & 182.44 & 60.80 & 0.01 &  \\
			\hline\hline
		\end{tabular}
		
	\end{table}

\begin{figure}
		\begin{center}
			\includegraphics[width=0.85\textwidth]{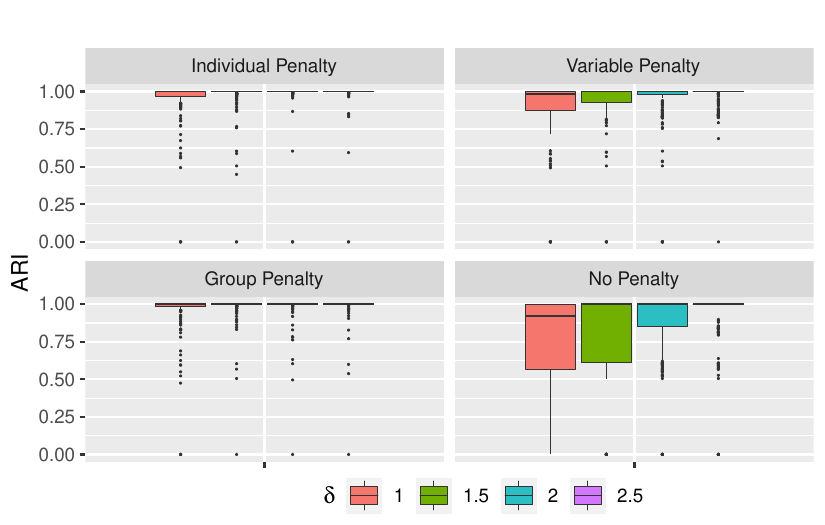}
			\caption{Boxplots of ARI for 200 datasets under different signal strengths for various methods.} \label{fig:simresult.signalStrength}
		\end{center}	
	\end{figure}

	Table~\ref{Table:Varianceevaluation} shows the summary of simulation results, and Figure~\ref{fig:simresult.signalStrength} shows the boxplots of ARI for 200 datasets under different signal strengths for various methods. The results show that stronger signals return better clustering results for all variable selection methods. In all cases, the group penalty method outperforms the variable and individual penalty methods in terms of $\MAE(\mhat)$ and the number of sensors removed falsely. The number of correctly removed sensors is close to the true value for all the cases and all the three different penalty methods. Similarly, the ARI plots show that the individual and group penalties are better than the variable penalty.

	\begin{table}
		\caption{Summary of simulation results under $\delta=1, 1.5, 2$, and $2.5$, with $n=200$, $\nsignal = 2$, and $\nnoisy = 16$ for all settings.}
		\label{Table:Varianceevaluation}
		\centering
		\begin{tabular}{c|c|c|c|c|c||c}\hline\hline
			\multirow{2}{*}{$\delta$ }& \multirow{2}{*}{Penalty} & \multirow{2}{*}{$\MAE(\mhat)$ }& Variable & Sensor removed & Sensor removed & $\MAE(\mhat)$    \\
			&  & &removed &  correctly& falsely & (no penalty) \\
			\hline
			
			\multirow{3}{*}{1} &
			Individual &
			0.36 & 49.98 & 15.88 & 0.04 &  \\
			&
			Variable &
			0.31 & 50.06 & 15.69 & 0.12 & 0.62 \\
			&
			Group &
			0.28 & 47.87 & 15.92 & 0.04 &  \\
			\hline
			\multirow{3}{*}{1.5} &
			Individual &
			0.26 & 49.95 & 15.91 & 0.02 &  \\
			&
			Variable &
			0.28 & 50.15 & 15.76 & 0.08 & 0.46 \\
			&
			Group &
			0.20 & 47.56 & 15.84 & 0.01 &  \\
			\hline
			\multirow{3}{*}{2} &
			Individual &
			0.26 & 49.98 & 15.91 & 0.04 &  \\
			&
			Variable &
			0.24 & 50.24 & 15.72 & 0.10 & 0.34 \\
			&
			Group &
			0.10 & 47.91 & 15.96 & 0.01 &  \\
			\hline
			\multirow{3}{*}{2.5} &
			Individual &
			0.20 & 49.92 & 15.89 & 0.02 &  \\
			&
			Variable &
			0.21 & 50.12 & 15.70 & 0.07 & 0.24 \\
			&
			Group &
			0.10 & 48.02 & 15.99 & 0.01 &  \\
			\hline\hline
		\end{tabular}
		
	\end{table}

	In summary, the group penalty performs the best in terms of ARI, $\MAE(\mhat)$, and the number of falsely removed sensors. Based on the simulation results, we recommend using the group penalty for the clustering of functional data with variable selection.

	\subsection{Comparison with Clustering without Variable Selections}
	Our focus is on functional clustering with variable selection. Because there are no existing methods for direct comparisons, we compare our methods with the method without variable selection. For clustering without variable selection, our approach is similar to \cite{jacques2014model}, but we use separate FPCA to handle a relatively large $p$.
	
	The results for clustering without variable selections are shown in the last columns of Tables~\ref{Table:SampleSize}, \ref{Table:Noisyevaluation}, and ~\ref{Table:Varianceevaluation}. In general, the $\MAE(\mhat)$ is much worse than those from variable selections. With larger sample sizes, $\MAE(\mhat)$ does become smaller for the method without variable selection, which shows it is necessary to consider variable selection especially when the sample size is small. When the number of noisy sensors increases, the $\MAE(\mhat)$ for the method without variable selection becomes worse, which shows it is important to consider variable selection when the number of noisy sensors is large.  The results show that stronger signals return better clustering results using the method without variable selection.
	
	The last panel of Figures~\ref{fig:simresult}, \ref{fig:simresult.noisySensor}, and \ref{fig:simresult.signalStrength} show the ARI plots for the method without variable selection. We observe similar patterns to those shown in the $\MAE(\mhat)$. In summary, both the results in the tables and ARI plots show that it is necessary to do variable selection when clustering functional data with noisy functional variables and when the sample size is small.

	\section{Application to System~B Data}\label{sec:data.application}
	In this section, we apply the proposed methods to the sensor data from System~B. The background of the application is introduced in Section~\ref{sec:introduction}. Based on the simulation results, we apply the proposed method with the group penalty to this engineering system sensor data, because the group penalty has the best performance.

We first re-scale the data to mean 0 and variance 1 before the analysis. That is, we first compute the sensor mean and standard deviation using the data from all curves from a sensor. Then, for each curve from the sensor, we subtract the curve by the sensor mean and then divide it by the sensor standard deviation. In the analysis, we use three FPCs to represent each sensor under the empirical rule proposed in Section~\ref{sec: hyperparameter}. In this case, we have $84\%$ of the sensors having more than $80\%$ of their variations explained by their first three FPCs. The hyper-parameters chosen by the adjusted BIC are $m = 4$, $\lambda = 11.2$, $\gamma = 2$ for the group penalty. The numbers of sensors removed and kept in the model are 23 and 20, respectively.

\begin{figure}
		
		\begin{center}
			\includegraphics[width=0.9\textwidth]{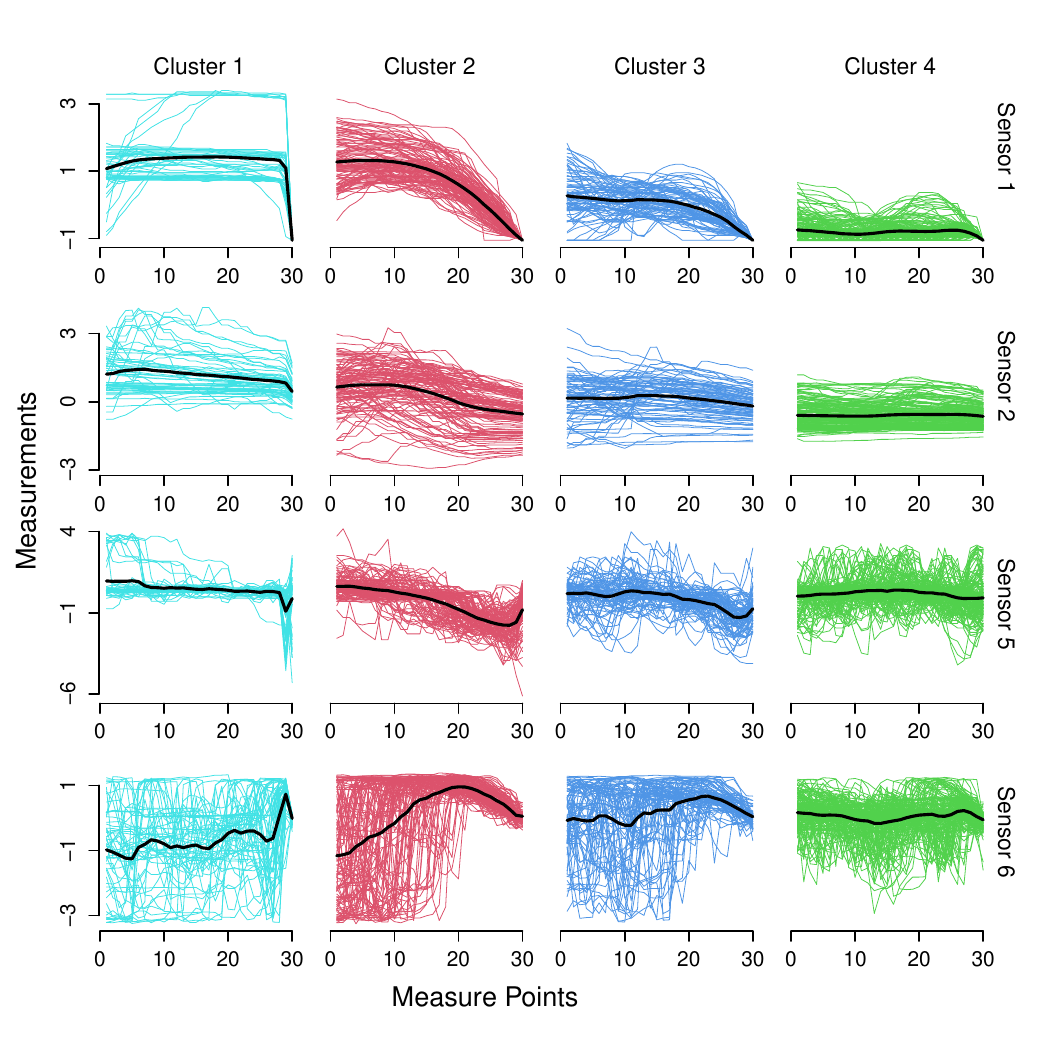}
			\caption{Clustering results for observations at four non-removed sensors in Engineer System~B using the group penalty. Four clusters are represented by different colors, and each colored line shows one observed functional curve from one sensor. The solid black lines represent cluster means.}
			\label{fig:selectedsensor}
		\end{center}	
	\end{figure}

\begin{figure}
		
		\begin{center}
			\includegraphics[width=0.9\textwidth]{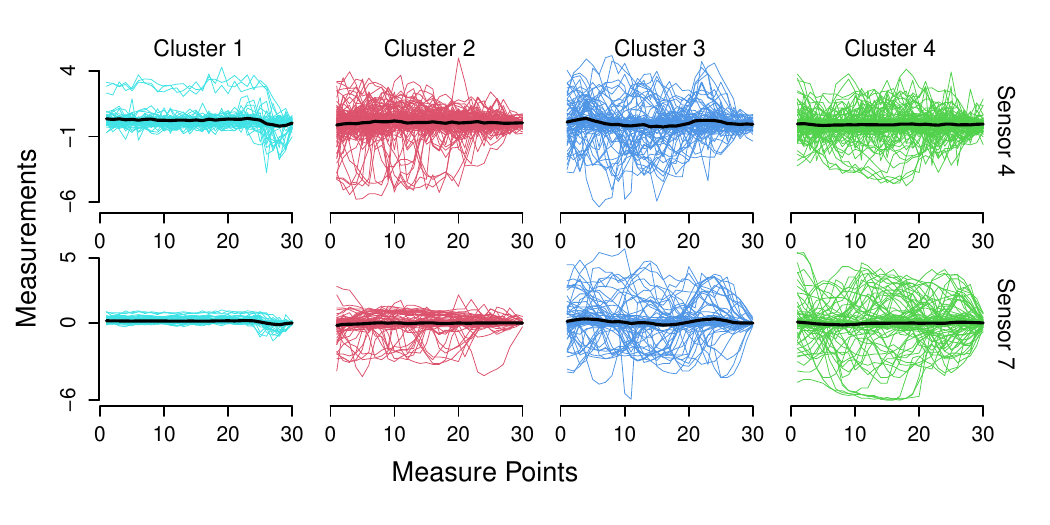}
			\caption{Clustering results for observations at two removed sensors in Engineer System~B using the group penalty. Four clusters are represented by different colors, and each colored line shows one observed functional curve from one sensor. The solid black lines represent cluster means.}
			\label{fig:removedsensor}
		\end{center}	
	\end{figure}

	We visualize the result from the group penalty method. Figure \ref{fig:selectedsensor} shows observations for a subset of four sensors that are not removed by the group penalty method, and Figure \ref{fig:removedsensor} shows observations for a subset of two sensors that are removed by the group penalty. The color represents four different clusters using the group penalty method. The solid black line in each panel shows the mean functional curve for each cluster.

	Clearly, the removed sensors in Figure \ref{fig:removedsensor} exhibit no group features, and the selected sensors in Figure \ref{fig:selectedsensor} show similar patterns within the same cluster. For example, the first row of Figure~\ref{fig:selectedsensor} shows an interesting pattern. The measurements of Sensor~1 decrease as time approaches the time point 30.  Cluster~1 corresponds to a pattern that Sensor~1 readings were high, and dropped drastically fast before time point 30.  Cluster~2 corresponds to an event pattern that Sensor~1 readings dropped gradually. Cluster~3 shows the readings dropped more slowly, and cluster~4 shows Sensor~1 readings kept at a lower value until time point 30. In the real application, our proposed methods provide engineers with an automatic way to extract information from large-scale sensor data.

	\section{Conclusions and Areas for Future Research}\label{sec:conclusionRemark}
	
	In this paper, we propose a clustering method for multivariate functional data with the unique feature of simultaneous variable selection. Three penalty terms are introduced for variable selection, namely, the individual, variable, and group penalties. All three penalties perform well in both clustering and variable selection in our simulation studies. However, the group penalty shows the best performance. Also, from the decision-making point of view, we used the EM method to optimize the utility function, which is the penalized negative log-likelihood. We suggest to use the modified adjusted BIC to select the number of clusters and the hyper-parameters in the clustering method.
	
    A sensor data application is considered in this paper, where sensors in the engineering system provide multivariate functional data. Selected sensors highlight important sensors contributing to the clustering of functional patterns. Our proposed methods can be used in the initial screening process of sensor scheduling and selection. The identification of important sensors and patterns through clustering provides an effective way to extract information from sensor data, which can be important for business owner of such data.

	Our method is applicable to applications other than engineering sensor systems. It can be applied to other studies with multivariate functional variables that have the need for clustering and variable selection. For example, a study on implantable body devices for identifying life-threatening events was presented in  \citet{VanLoNgKing2004}, where only a K-means clustering is employed. Our method could enhance the clustering with the incorporation of variable selection. In addition, indoor environmental safety issues occur more frequently because of the increased usage of home appliances. Usually multiple devices are deployed to monitor possible emergencies such as fire and smoke appearances (\citealp{WuDerek2007}). Our method could be applied to recognize various risks at home.

	As for future research, the robustness of clustering methods has been of interest in literature (e.g., \citealp{Tupperetal2018} and \citealp{ParkSimpson2019}). It will be interesting to study the robustness of functional clustering methods under the setting of variable selection, which is non-trivial. In another direction, misalignment of the functional data can be challenging in clustering (e.g., \citealp{SlaetsClaeskensHubert2012}). Although our sensor data are well aligned, it will be interesting to consider functional clustering with variable selection under functional data with misalignment.
	
In some applications, the sampling frequencies may vary among different sensors. Sampling frequencies may influence the clustering results through the estimated values of the truncated coefficients of the Karhunen-Lo\`{e}ve expansion in FPCA. When some sensors do not have enough samples, the clustering results may change. Investigating and solving the inadequate sample problem is an interesting future research topic in functional data clustering. Another potential future research direction is selecting the number of clusters and useful sensors using methods other than the adjusted BIC. Previously, \citet{huang2002varying} and \citet{hurvich1998smoothing} developed model selection criteria for selecting the smoothing hyper-parameters in estimating functional curves. Those two papers can bring insights into improving the adjusted BIC selection method in our case. Thus, it is worthy to further explore the improvement opportunity in the future.

\section*{Supplementary Material}

The following supplementary materials are available online.

\begin{description}
\item[Code and data:] Processed data, and R code for data simulations (zip file).
	
\end{description}

	\section*{Acknowledgments}

The authors thank the Editor-in-Chief, a Senior Editor, an Associate Editor, and three anonymous referees for providing helpful comments and suggestions that significantly improved this paper. The authors acknowledge Advanced Research Computing at Virginia Tech for providing computational resources.

	\section*{Funding}
	The research by Min and Hong was partially supported by National Science Foundation Grant CMMI-1904165 to Virginia Tech. The work by Hong was partially supported by the Virginia Tech College of Science Research Equipment Fund.

	\begin{appendix}
		\section{Some Derivations}
		
		\subsection{Mean Component Estimates}\label{Appendix:UnconstrianedMu}

		Let $\widetilde{\mu}_{kj}$ be the minimizer for \eqref{equ: likelihood_fun}, when there is no constraint on the likelihood, and $\widetilde{\mu}_{kj}$ satisfies
		$$\left.\frac{\partial l(\thetavec) }{\partial \mu_{kj}} \right| _ {\mu_{kj} = \widetilde{\mu}_{kj}} = 0. $$
		The partial derivative of the likelihood is then
		
		\begin{align}
			\frac{\partial l(\thetavec)}{\partial \muvec_{k}} &= \frac{\partial }{\partial  \muvec_{k}} \sum_{i=1}^{n}-\delta_{ik} \left\{ \log(\pi_{k})+\log \left[ f_{k}(\bvec_{i}; \muvec_{k}, \Sigma) \right] \right\} \nonumber \\
			&= - \sum_{i=1}^{n} \delta_{ik} \Sigma^{-1} (\bvec_{i} - \muvec_{k}).\label{eqn:derivedPDwrtMu}
		\end{align}
		With a non-singular covariance matrix $\Sigma$, we obtain the estimator of $\muvec_{k}$ by setting \eqref{eqn:derivedPDwrtMu} to zero and multiplying $\Sigma$ on both sides of the equation. That is,
		$\sum_{i=1}^{n} \delta_{ik}(\bvec_{i} - \muvec_{k})  = \zerovec.$
		Then the estimate for the mean component from cluster $k$ is
		$$\widetilde{\muvec}_{k}= \sum_{i = 1}^{n} \delta_{ik} \bvec_{i}/ \sum_{i=1}^{n} \delta_{ik},$$
		while the indicator $\delta_{ik}$ is replaced by its estimate $\tauhat_{ik}$.
		
		\subsection{Covariance Matrix in the EM Algorithm} \label{Appendix:CovarianceMatrixCal}
		Similar to the mean component estimate calculation, $\sigmahat^{2}_{j}$'s are calculated by setting the partial derivative of the penalized log-likelihood function with respect to $\Sigma$ to be zero.
		We have,
		\begin{align}
			\frac{\partial \lpen(\thetavec)}{\partial \sigma_j^2} &= \frac{\partial }{\partial \sigma_j^2} \left\{ \sum_{i=1}^{n}\sum_{k=1}^{m} -\delta_{ik}  \left\{ \log(\pi_{k})+\log \left[ f_{k}(\bvec_{i}; \muvec_{k}, \Sigma) \right] \right\} +p_{\lambda}(\thetavec) \right\} \nonumber \\
			&= \sum_{i=1}^{n}\sum_{k=1}^{m} \frac{\delta_{ik}}{2}\left[ \trace\left(\Sigma^{-1}\frac{\partial\Sigma}{\sigma_j^2}\right)-\trace\left(\Sigma^{-1}A_{ik}\Sigma^{-1}
			\frac{\partial\Sigma}{\partial\sigma_j^2}\right)\right]
			\nonumber \\
			&= \sum_{i=1}^{n}\sum_{k=1}^{m}  \frac{\delta_{ik}}{2} \left[\frac{1}{\sigma_j^2} -  \frac{(b_{ij}-\muhat_{kj})^2}{\sigma_j^4} \right], \label{eqn:derivedPDwrtSigma}
		\end{align}
		where $A_{ik}=(\bvec_{i} - \muvec_{k}) (\bvec_{i} - \muvec_{k})'$. According to the definition, $\sum_{i=1}^{n}\sum_{k=1}^{m} \delta_{ik}  = n$ is the total number of observations. Setting \eqref{eqn:derivedPDwrtSigma} to $\zerovec$, we have the $j$th diagonal element  in $\Sigma$ calculated as
		$$\sigmahat^{2}_{j} = \sum_{k=1}^{m} \sum_{i = 1}^{n} \delta_{ik} (b_{ij} - \muhat_{kj})^{2}/n,$$
		while the indicator $\delta_{ik}$ is replaced by its estimate $\tauhat_{ik}$.

	\end{appendix}


\end{document}